\documentclass[a4paper,lnbip]{svmultln}
\usepackage{graphicx}
\usepackage{tabularx}
\usepackage{float}

\graphicspath{{img/}}

\usepackage[%
draft,% 
usepdfnote,%
showoldrev,%
marknewrev%
]{todos} 
\usepackage{color}
\definecolor{AliceBlue}{rgb}{0.94,0.97,1.00}
\definecolor{AntiqueWhite1}{rgb}{1.00,0.94,0.86}
\definecolor{AntiqueWhite2}{rgb}{0.93,0.87,0.80}
\definecolor{AntiqueWhite3}{rgb}{0.80,0.75,0.69}
\definecolor{AntiqueWhite4}{rgb}{0.55,0.51,0.47}
\definecolor{AntiqueWhite}{rgb}{0.98,0.92,0.84}
\definecolor{BlanchedAlmond}{rgb}{1.00,0.92,0.80}
\definecolor{BlueViolet}{rgb}{0.54,0.17,0.89}
\definecolor{CadetBlue1}{rgb}{0.60,0.96,1.00}
\definecolor{CadetBlue2}{rgb}{0.56,0.90,0.93}
\definecolor{CadetBlue3}{rgb}{0.48,0.77,0.80}
\definecolor{CadetBlue4}{rgb}{0.33,0.53,0.55}
\definecolor{CadetBlue}{rgb}{0.37,0.62,0.63}
\definecolor{CornflowerBlue}{rgb}{0.39,0.58,0.93}
\definecolor{DarkBlue}{rgb}{0.00,0.00,0.55}
\definecolor{DarkCyan}{rgb}{0.00,0.55,0.55}
\definecolor{DarkGoldenrod1}{rgb}{1.00,0.73,0.06}
\definecolor{DarkGoldenrod2}{rgb}{0.93,0.68,0.05}
\definecolor{DarkGoldenrod3}{rgb}{0.80,0.58,0.05}
\definecolor{DarkGoldenrod4}{rgb}{0.55,0.40,0.03}
\definecolor{DarkGoldenrod}{rgb}{0.72,0.53,0.04}
\definecolor{DarkGray}{rgb}{0.66,0.66,0.66}
\definecolor{DarkGreen}{rgb}{0.00,0.39,0.00}
\definecolor{DarkGrey}{rgb}{0.66,0.66,0.66}
\definecolor{DarkKhaki}{rgb}{0.74,0.72,0.42}
\definecolor{DarkMagenta}{rgb}{0.55,0.00,0.55}
\definecolor{DarkOliveGreen1}{rgb}{0.79,1.00,0.44}
\definecolor{DarkOliveGreen2}{rgb}{0.74,0.93,0.41}
\definecolor{DarkOliveGreen3}{rgb}{0.64,0.80,0.35}
\definecolor{DarkOliveGreen4}{rgb}{0.43,0.55,0.24}
\definecolor{DarkOliveGreen}{rgb}{0.33,0.42,0.18}
\definecolor{DarkOrange1}{rgb}{1.00,0.50,0.00}
\definecolor{DarkOrange2}{rgb}{0.93,0.46,0.00}
\definecolor{DarkOrange3}{rgb}{0.80,0.40,0.00}
\definecolor{DarkOrange4}{rgb}{0.55,0.27,0.00}
\definecolor{DarkOrange}{rgb}{1.00,0.55,0.00}
\definecolor{DarkOrchid1}{rgb}{0.75,0.24,1.00}
\definecolor{DarkOrchid2}{rgb}{0.70,0.23,0.93}
\definecolor{DarkOrchid3}{rgb}{0.60,0.20,0.80}
\definecolor{DarkOrchid4}{rgb}{0.41,0.13,0.55}
\definecolor{DarkOrchid}{rgb}{0.60,0.20,0.80}
\definecolor{DarkRed}{rgb}{0.55,0.00,0.00}
\definecolor{DarkSalmon}{rgb}{0.91,0.59,0.48}
\definecolor{DarkSeaGreen1}{rgb}{0.76,1.00,0.76}
\definecolor{DarkSeaGreen2}{rgb}{0.71,0.93,0.71}
\definecolor{DarkSeaGreen3}{rgb}{0.61,0.80,0.61}
\definecolor{DarkSeaGreen4}{rgb}{0.41,0.55,0.41}
\definecolor{DarkSeaGreen}{rgb}{0.56,0.74,0.56}
\definecolor{DarkSlateBlue}{rgb}{0.28,0.24,0.55}
\definecolor{DarkSlateGray1}{rgb}{0.59,1.00,1.00}
\definecolor{DarkSlateGray2}{rgb}{0.55,0.93,0.93}
\definecolor{DarkSlateGray3}{rgb}{0.47,0.80,0.80}
\definecolor{DarkSlateGray4}{rgb}{0.32,0.55,0.55}
\definecolor{DarkSlateGray}{rgb}{0.18,0.31,0.31}
\definecolor{DarkSlateGrey}{rgb}{0.18,0.31,0.31}
\definecolor{DarkTurquoise}{rgb}{0.00,0.81,0.82}
\definecolor{DarkViolet}{rgb}{0.58,0.00,0.83}
\definecolor{DeepPink1}{rgb}{1.00,0.08,0.58}
\definecolor{DeepPink2}{rgb}{0.93,0.07,0.54}
\definecolor{DeepPink3}{rgb}{0.80,0.06,0.46}
\definecolor{DeepPink4}{rgb}{0.55,0.04,0.31}
\definecolor{DeepPink}{rgb}{1.00,0.08,0.58}
\definecolor{DeepSkyBlue1}{rgb}{0.00,0.75,1.00}
\definecolor{DeepSkyBlue2}{rgb}{0.00,0.70,0.93}
\definecolor{DeepSkyBlue3}{rgb}{0.00,0.60,0.80}
\definecolor{DeepSkyBlue4}{rgb}{0.00,0.41,0.55}
\definecolor{DeepSkyBlue}{rgb}{0.00,0.75,1.00}
\definecolor{DimGray}{rgb}{0.41,0.41,0.41}
\definecolor{DimGrey}{rgb}{0.41,0.41,0.41}
\definecolor{DodgerBlue1}{rgb}{0.12,0.56,1.00}
\definecolor{DodgerBlue2}{rgb}{0.11,0.53,0.93}
\definecolor{DodgerBlue3}{rgb}{0.09,0.45,0.80}
\definecolor{DodgerBlue4}{rgb}{0.06,0.31,0.55}
\definecolor{DodgerBlue}{rgb}{0.12,0.56,1.00}
\definecolor{FloralWhite}{rgb}{1.00,0.98,0.94}
\definecolor{ForestGreen}{rgb}{0.13,0.55,0.13}
\definecolor{GhostWhite}{rgb}{0.97,0.97,1.00}
\definecolor{GreenYellow}{rgb}{0.68,1.00,0.18}
\definecolor{HotPink1}{rgb}{1.00,0.43,0.71}
\definecolor{HotPink2}{rgb}{0.93,0.42,0.65}
\definecolor{HotPink3}{rgb}{0.80,0.38,0.56}
\definecolor{HotPink4}{rgb}{0.55,0.23,0.38}
\definecolor{HotPink}{rgb}{1.00,0.41,0.71}
\definecolor{IndianRed1}{rgb}{1.00,0.42,0.42}
\definecolor{IndianRed2}{rgb}{0.93,0.39,0.39}
\definecolor{IndianRed3}{rgb}{0.80,0.33,0.33}
\definecolor{IndianRed4}{rgb}{0.55,0.23,0.23}
\definecolor{IndianRed}{rgb}{0.80,0.36,0.36}
\definecolor{LavenderBlush1}{rgb}{1.00,0.94,0.96}
\definecolor{LavenderBlush2}{rgb}{0.93,0.88,0.90}
\definecolor{LavenderBlush3}{rgb}{0.80,0.76,0.77}
\definecolor{LavenderBlush4}{rgb}{0.55,0.51,0.53}
\definecolor{LavenderBlush}{rgb}{1.00,0.94,0.96}
\definecolor{LawnGreen}{rgb}{0.49,0.99,0.00}
\definecolor{LemonChiffon1}{rgb}{1.00,0.98,0.80}
\definecolor{LemonChiffon2}{rgb}{0.93,0.91,0.75}
\definecolor{LemonChiffon3}{rgb}{0.80,0.79,0.65}
\definecolor{LemonChiffon4}{rgb}{0.55,0.54,0.44}
\definecolor{LemonChiffon}{rgb}{1.00,0.98,0.80}
\definecolor{LightBlue1}{rgb}{0.75,0.94,1.00}
\definecolor{LightBlue2}{rgb}{0.70,0.87,0.93}
\definecolor{LightBlue3}{rgb}{0.60,0.75,0.80}
\definecolor{LightBlue4}{rgb}{0.41,0.51,0.55}
\definecolor{LightBlue}{rgb}{0.68,0.85,0.90}
\definecolor{LightCoral}{rgb}{0.94,0.50,0.50}
\definecolor{LightCyan1}{rgb}{0.88,1.00,1.00}
\definecolor{LightCyan2}{rgb}{0.82,0.93,0.93}
\definecolor{LightCyan3}{rgb}{0.71,0.80,0.80}
\definecolor{LightCyan4}{rgb}{0.48,0.55,0.55}
\definecolor{LightCyan}{rgb}{0.88,1.00,1.00}
\definecolor{LightGoldenrod1}{rgb}{1.00,0.93,0.55}
\definecolor{LightGoldenrod2}{rgb}{0.93,0.86,0.51}
\definecolor{LightGoldenrod3}{rgb}{0.80,0.75,0.44}
\definecolor{LightGoldenrod4}{rgb}{0.55,0.51,0.30}
\definecolor{LightGoldenrodYellow}{rgb}{0.98,0.98,0.82}
\definecolor{LightGoldenrod}{rgb}{0.93,0.87,0.51}
\definecolor{LightGray}{rgb}{0.83,0.83,0.83}
\definecolor{LightGreen}{rgb}{0.56,0.93,0.56}
\definecolor{LightGrey}{rgb}{0.83,0.83,0.83}
\definecolor{LightPink1}{rgb}{1.00,0.68,0.73}
\definecolor{LightPink2}{rgb}{0.93,0.64,0.68}
\definecolor{LightPink3}{rgb}{0.80,0.55,0.58}
\definecolor{LightPink4}{rgb}{0.55,0.37,0.40}
\definecolor{LightPink}{rgb}{1.00,0.71,0.76}
\definecolor{LightSalmon1}{rgb}{1.00,0.63,0.48}
\definecolor{LightSalmon2}{rgb}{0.93,0.58,0.45}
\definecolor{LightSalmon3}{rgb}{0.80,0.51,0.38}
\definecolor{LightSalmon4}{rgb}{0.55,0.34,0.26}
\definecolor{LightSalmon}{rgb}{1.00,0.63,0.48}
\definecolor{LightSeaGreen}{rgb}{0.13,0.70,0.67}
\definecolor{LightSkyBlue1}{rgb}{0.69,0.89,1.00}
\definecolor{LightSkyBlue2}{rgb}{0.64,0.83,0.93}
\definecolor{LightSkyBlue3}{rgb}{0.55,0.71,0.80}
\definecolor{LightSkyBlue4}{rgb}{0.38,0.48,0.55}
\definecolor{LightSkyBlue}{rgb}{0.53,0.81,0.98}
\definecolor{LightSlateBlue}{rgb}{0.52,0.44,1.00}
\definecolor{LightSlateGray}{rgb}{0.47,0.53,0.60}
\definecolor{LightSlateGrey}{rgb}{0.47,0.53,0.60}
\definecolor{LightSteelBlue1}{rgb}{0.79,0.88,1.00}
\definecolor{LightSteelBlue2}{rgb}{0.74,0.82,0.93}
\definecolor{LightSteelBlue3}{rgb}{0.64,0.71,0.80}
\definecolor{LightSteelBlue4}{rgb}{0.43,0.48,0.55}
\definecolor{LightSteelBlue}{rgb}{0.69,0.77,0.87}
\definecolor{LightYellow1}{rgb}{1.00,1.00,0.88}
\definecolor{LightYellow2}{rgb}{0.93,0.93,0.82}
\definecolor{LightYellow3}{rgb}{0.80,0.80,0.71}
\definecolor{LightYellow4}{rgb}{0.55,0.55,0.48}
\definecolor{LightYellow}{rgb}{1.00,1.00,0.88}
\definecolor{LimeGreen}{rgb}{0.20,0.80,0.20}
\definecolor{MediumAquamarine}{rgb}{0.40,0.80,0.67}
\definecolor{MediumBlue}{rgb}{0.00,0.00,0.80}
\definecolor{MediumOrchid1}{rgb}{0.88,0.40,1.00}
\definecolor{MediumOrchid2}{rgb}{0.82,0.37,0.93}
\definecolor{MediumOrchid3}{rgb}{0.71,0.32,0.80}
\definecolor{MediumOrchid4}{rgb}{0.48,0.22,0.55}
\definecolor{MediumOrchid}{rgb}{0.73,0.33,0.83}
\definecolor{MediumPurple1}{rgb}{0.67,0.51,1.00}
\definecolor{MediumPurple2}{rgb}{0.62,0.47,0.93}
\definecolor{MediumPurple3}{rgb}{0.54,0.41,0.80}
\definecolor{MediumPurple4}{rgb}{0.36,0.28,0.55}
\definecolor{MediumPurple}{rgb}{0.58,0.44,0.86}
\definecolor{MediumSeaGreen}{rgb}{0.24,0.70,0.44}
\definecolor{MediumSlateBlue}{rgb}{0.48,0.41,0.93}
\definecolor{MediumSpringGreen}{rgb}{0.00,0.98,0.60}
\definecolor{MediumTurquoise}{rgb}{0.28,0.82,0.80}
\definecolor{MediumVioletRed}{rgb}{0.78,0.08,0.52}
\definecolor{MidnightBlue}{rgb}{0.10,0.10,0.44}
\definecolor{MintCream}{rgb}{0.96,1.00,0.98}
\definecolor{MistyRose1}{rgb}{1.00,0.89,0.88}
\definecolor{MistyRose2}{rgb}{0.93,0.84,0.82}
\definecolor{MistyRose3}{rgb}{0.80,0.72,0.71}
\definecolor{MistyRose4}{rgb}{0.55,0.49,0.48}
\definecolor{MistyRose}{rgb}{1.00,0.89,0.88}
\definecolor{NavajoWhite1}{rgb}{1.00,0.87,0.68}
\definecolor{NavajoWhite2}{rgb}{0.93,0.81,0.63}
\definecolor{NavajoWhite3}{rgb}{0.80,0.70,0.55}
\definecolor{NavajoWhite4}{rgb}{0.55,0.47,0.37}
\definecolor{NavajoWhite}{rgb}{1.00,0.87,0.68}
\definecolor{NavyBlue}{rgb}{0.00,0.00,0.50}
\definecolor{OldLace}{rgb}{0.99,0.96,0.90}
\definecolor{OliveDrab1}{rgb}{0.75,1.00,0.24}
\definecolor{OliveDrab2}{rgb}{0.70,0.93,0.23}
\definecolor{OliveDrab3}{rgb}{0.60,0.80,0.20}
\definecolor{OliveDrab4}{rgb}{0.41,0.55,0.13}
\definecolor{OliveDrab}{rgb}{0.42,0.56,0.14}
\definecolor{OrangeRed1}{rgb}{1.00,0.27,0.00}
\definecolor{OrangeRed2}{rgb}{0.93,0.25,0.00}
\definecolor{OrangeRed3}{rgb}{0.80,0.22,0.00}
\definecolor{OrangeRed4}{rgb}{0.55,0.15,0.00}
\definecolor{OrangeRed}{rgb}{1.00,0.27,0.00}
\definecolor{PaleGoldenrod}{rgb}{0.93,0.91,0.67}
\definecolor{PaleGreen1}{rgb}{0.60,1.00,0.60}
\definecolor{PaleGreen2}{rgb}{0.56,0.93,0.56}
\definecolor{PaleGreen3}{rgb}{0.49,0.80,0.49}
\definecolor{PaleGreen4}{rgb}{0.33,0.55,0.33}
\definecolor{PaleGreen}{rgb}{0.60,0.98,0.60}
\definecolor{PaleTurquoise1}{rgb}{0.73,1.00,1.00}
\definecolor{PaleTurquoise2}{rgb}{0.68,0.93,0.93}
\definecolor{PaleTurquoise3}{rgb}{0.59,0.80,0.80}
\definecolor{PaleTurquoise4}{rgb}{0.40,0.55,0.55}
\definecolor{PaleTurquoise}{rgb}{0.69,0.93,0.93}
\definecolor{PaleVioletRed1}{rgb}{1.00,0.51,0.67}
\definecolor{PaleVioletRed2}{rgb}{0.93,0.47,0.62}
\definecolor{PaleVioletRed3}{rgb}{0.80,0.41,0.54}
\definecolor{PaleVioletRed4}{rgb}{0.55,0.28,0.36}
\definecolor{PaleVioletRed}{rgb}{0.86,0.44,0.58}
\definecolor{PapayaWhip}{rgb}{1.00,0.94,0.84}
\definecolor{PeachPuff1}{rgb}{1.00,0.85,0.73}
\definecolor{PeachPuff2}{rgb}{0.93,0.80,0.68}
\definecolor{PeachPuff3}{rgb}{0.80,0.69,0.58}
\definecolor{PeachPuff4}{rgb}{0.55,0.47,0.40}
\definecolor{PeachPuff}{rgb}{1.00,0.85,0.73}
\definecolor{PowderBlue}{rgb}{0.69,0.88,0.90}
\definecolor{RosyBrown1}{rgb}{1.00,0.76,0.76}
\definecolor{RosyBrown2}{rgb}{0.93,0.71,0.71}
\definecolor{RosyBrown3}{rgb}{0.80,0.61,0.61}
\definecolor{RosyBrown4}{rgb}{0.55,0.41,0.41}
\definecolor{RosyBrown}{rgb}{0.74,0.56,0.56}
\definecolor{RoyalBlue1}{rgb}{0.28,0.46,1.00}
\definecolor{RoyalBlue2}{rgb}{0.26,0.43,0.93}
\definecolor{RoyalBlue3}{rgb}{0.23,0.37,0.80}
\definecolor{RoyalBlue4}{rgb}{0.15,0.25,0.55}
\definecolor{RoyalBlue}{rgb}{0.25,0.41,0.88}
\definecolor{SaddleBrown}{rgb}{0.55,0.27,0.07}
\definecolor{SandyBrown}{rgb}{0.96,0.64,0.38}
\definecolor{SeaGreen1}{rgb}{0.33,1.00,0.62}
\definecolor{SeaGreen2}{rgb}{0.31,0.93,0.58}
\definecolor{SeaGreen3}{rgb}{0.26,0.80,0.50}
\definecolor{SeaGreen4}{rgb}{0.18,0.55,0.34}
\definecolor{SeaGreen}{rgb}{0.18,0.55,0.34}
\definecolor{SkyBlue1}{rgb}{0.53,0.81,1.00}
\definecolor{SkyBlue2}{rgb}{0.49,0.75,0.93}
\definecolor{SkyBlue3}{rgb}{0.42,0.65,0.80}
\definecolor{SkyBlue4}{rgb}{0.29,0.44,0.55}
\definecolor{SkyBlue}{rgb}{0.53,0.81,0.92}
\definecolor{SlateBlue1}{rgb}{0.51,0.44,1.00}
\definecolor{SlateBlue2}{rgb}{0.48,0.40,0.93}
\definecolor{SlateBlue3}{rgb}{0.41,0.35,0.80}
\definecolor{SlateBlue4}{rgb}{0.28,0.24,0.55}
\definecolor{SlateBlue}{rgb}{0.42,0.35,0.80}
\definecolor{SlateGray1}{rgb}{0.78,0.89,1.00}
\definecolor{SlateGray2}{rgb}{0.73,0.83,0.93}
\definecolor{SlateGray3}{rgb}{0.62,0.71,0.80}
\definecolor{SlateGray4}{rgb}{0.42,0.48,0.55}
\definecolor{SlateGray}{rgb}{0.44,0.50,0.56}
\definecolor{SlateGrey}{rgb}{0.44,0.50,0.56}
\definecolor{SpringGreen1}{rgb}{0.00,1.00,0.50}
\definecolor{SpringGreen2}{rgb}{0.00,0.93,0.46}
\definecolor{SpringGreen3}{rgb}{0.00,0.80,0.40}
\definecolor{SpringGreen4}{rgb}{0.00,0.55,0.27}
\definecolor{SpringGreen}{rgb}{0.00,1.00,0.50}
\definecolor{SteelBlue1}{rgb}{0.39,0.72,1.00}
\definecolor{SteelBlue2}{rgb}{0.36,0.67,0.93}
\definecolor{SteelBlue3}{rgb}{0.31,0.58,0.80}
\definecolor{SteelBlue4}{rgb}{0.21,0.39,0.55}
\definecolor{SteelBlue}{rgb}{0.27,0.51,0.71}
\definecolor{VioletRed1}{rgb}{1.00,0.24,0.59}
\definecolor{VioletRed2}{rgb}{0.93,0.23,0.55}
\definecolor{VioletRed3}{rgb}{0.80,0.20,0.47}
\definecolor{VioletRed4}{rgb}{0.55,0.13,0.32}
\definecolor{VioletRed}{rgb}{0.82,0.13,0.56}
\definecolor{WhiteSmoke}{rgb}{0.96,0.96,0.96}
\definecolor{YellowGreen}{rgb}{0.60,0.80,0.20}
\definecolor{aliceblue}{rgb}{0.94,0.97,1.00}
\definecolor{antiquewhite}{rgb}{0.98,0.92,0.84}
\definecolor{aquamarine1}{rgb}{0.50,1.00,0.83}
\definecolor{aquamarine2}{rgb}{0.46,0.93,0.78}
\definecolor{aquamarine3}{rgb}{0.40,0.80,0.67}
\definecolor{aquamarine4}{rgb}{0.27,0.55,0.45}
\definecolor{aquamarine}{rgb}{0.50,1.00,0.83}
\definecolor{azure1}{rgb}{0.94,1.00,1.00}
\definecolor{azure2}{rgb}{0.88,0.93,0.93}
\definecolor{azure3}{rgb}{0.76,0.80,0.80}
\definecolor{azure4}{rgb}{0.51,0.55,0.55}
\definecolor{azure}{rgb}{0.94,1.00,1.00}
\definecolor{beige}{rgb}{0.96,0.96,0.86}
\definecolor{bisque1}{rgb}{1.00,0.89,0.77}
\definecolor{bisque2}{rgb}{0.93,0.84,0.72}
\definecolor{bisque3}{rgb}{0.80,0.72,0.62}
\definecolor{bisque4}{rgb}{0.55,0.49,0.42}
\definecolor{bisque}{rgb}{1.00,0.89,0.77}
\definecolor{black}{rgb}{0.00,0.00,0.00}
\definecolor{blanchedalmond}{rgb}{1.00,0.92,0.80}
\definecolor{blue1}{rgb}{0.00,0.00,1.00}
\definecolor{blue2}{rgb}{0.00,0.00,0.93}
\definecolor{blue3}{rgb}{0.00,0.00,0.80}
\definecolor{blue4}{rgb}{0.00,0.00,0.55}
\definecolor{blueviolet}{rgb}{0.54,0.17,0.89}
\definecolor{blue}{rgb}{0.00,0.00,1.00}
\definecolor{brown1}{rgb}{1.00,0.25,0.25}
\definecolor{brown2}{rgb}{0.93,0.23,0.23}
\definecolor{brown3}{rgb}{0.80,0.20,0.20}
\definecolor{brown4}{rgb}{0.55,0.14,0.14}
\definecolor{brown}{rgb}{0.65,0.16,0.16}
\definecolor{burlywood1}{rgb}{1.00,0.83,0.61}
\definecolor{burlywood2}{rgb}{0.93,0.77,0.57}
\definecolor{burlywood3}{rgb}{0.80,0.67,0.49}
\definecolor{burlywood4}{rgb}{0.55,0.45,0.33}
\definecolor{burlywood}{rgb}{0.87,0.72,0.53}
\definecolor{cadetblue}{rgb}{0.37,0.62,0.63}
\definecolor{chartreuse1}{rgb}{0.50,1.00,0.00}
\definecolor{chartreuse2}{rgb}{0.46,0.93,0.00}
\definecolor{chartreuse3}{rgb}{0.40,0.80,0.00}
\definecolor{chartreuse4}{rgb}{0.27,0.55,0.00}
\definecolor{chartreuse}{rgb}{0.50,1.00,0.00}
\definecolor{chocolate1}{rgb}{1.00,0.50,0.14}
\definecolor{chocolate2}{rgb}{0.93,0.46,0.13}
\definecolor{chocolate3}{rgb}{0.80,0.40,0.11}
\definecolor{chocolate4}{rgb}{0.55,0.27,0.07}
\definecolor{chocolate}{rgb}{0.82,0.41,0.12}
\definecolor{coral1}{rgb}{1.00,0.45,0.34}
\definecolor{coral2}{rgb}{0.93,0.42,0.31}
\definecolor{coral3}{rgb}{0.80,0.36,0.27}
\definecolor{coral4}{rgb}{0.55,0.24,0.18}
\definecolor{coral}{rgb}{1.00,0.50,0.31}
\definecolor{cornflowerblue}{rgb}{0.39,0.58,0.93}
\definecolor{cornsilk1}{rgb}{1.00,0.97,0.86}
\definecolor{cornsilk2}{rgb}{0.93,0.91,0.80}
\definecolor{cornsilk3}{rgb}{0.80,0.78,0.69}
\definecolor{cornsilk4}{rgb}{0.55,0.53,0.47}
\definecolor{cornsilk}{rgb}{1.00,0.97,0.86}
\definecolor{cyan1}{rgb}{0.00,1.00,1.00}
\definecolor{cyan2}{rgb}{0.00,0.93,0.93}
\definecolor{cyan3}{rgb}{0.00,0.80,0.80}
\definecolor{cyan4}{rgb}{0.00,0.55,0.55}
\definecolor{cyan}{rgb}{0.00,1.00,1.00}
\definecolor{darkblue}{rgb}{0.00,0.00,0.55}
\definecolor{darkcyan}{rgb}{0.00,0.55,0.55}
\definecolor{darkgoldenrod}{rgb}{0.72,0.53,0.04}
\definecolor{darkgray}{rgb}{0.66,0.66,0.66}
\definecolor{darkgreen}{rgb}{0.00,0.39,0.00}
\definecolor{darkgrey}{rgb}{0.66,0.66,0.66}
\definecolor{darkkhaki}{rgb}{0.74,0.72,0.42}
\definecolor{darkmagenta}{rgb}{0.55,0.00,0.55}
\definecolor{darkolive}{rgb}{0.33,0.42,0.18}
\definecolor{darkorange}{rgb}{1.00,0.55,0.00}
\definecolor{darkorchid}{rgb}{0.60,0.20,0.80}
\definecolor{darkred}{rgb}{0.55,0.00,0.00}
\definecolor{darksalmon}{rgb}{0.91,0.59,0.48}
\definecolor{darksea}{rgb}{0.56,0.74,0.56}
\definecolor{darkslate}{rgb}{0.18,0.31,0.31}
\definecolor{darkslate}{rgb}{0.18,0.31,0.31}
\definecolor{darkslate}{rgb}{0.28,0.24,0.55}
\definecolor{darkturquoise}{rgb}{0.00,0.81,0.82}
\definecolor{darkviolet}{rgb}{0.58,0.00,0.83}
\definecolor{deeppink}{rgb}{1.00,0.08,0.58}
\definecolor{deepsky}{rgb}{0.00,0.75,1.00}
\definecolor{dimgray}{rgb}{0.41,0.41,0.41}
\definecolor{dimgrey}{rgb}{0.41,0.41,0.41}
\definecolor{dodgerblue}{rgb}{0.12,0.56,1.00}
\definecolor{firebrick1}{rgb}{1.00,0.19,0.19}
\definecolor{firebrick2}{rgb}{0.93,0.17,0.17}
\definecolor{firebrick3}{rgb}{0.80,0.15,0.15}
\definecolor{firebrick4}{rgb}{0.55,0.10,0.10}
\definecolor{firebrick}{rgb}{0.70,0.13,0.13}
\definecolor{floralwhite}{rgb}{1.00,0.98,0.94}
\definecolor{forestgreen}{rgb}{0.13,0.55,0.13}
\definecolor{gainsboro}{rgb}{0.86,0.86,0.86}
\definecolor{ghostwhite}{rgb}{0.97,0.97,1.00}
\definecolor{gold1}{rgb}{1.00,0.84,0.00}
\definecolor{gold2}{rgb}{0.93,0.79,0.00}
\definecolor{gold3}{rgb}{0.80,0.68,0.00}
\definecolor{gold4}{rgb}{0.55,0.46,0.00}
\definecolor{goldenrod1}{rgb}{1.00,0.76,0.15}
\definecolor{goldenrod2}{rgb}{0.93,0.71,0.13}
\definecolor{goldenrod3}{rgb}{0.80,0.61,0.11}
\definecolor{goldenrod4}{rgb}{0.55,0.41,0.08}
\definecolor{goldenrod}{rgb}{0.85,0.65,0.13}
\definecolor{gold}{rgb}{1.00,0.84,0.00}
\definecolor{gray0}{rgb}{0.00,0.00,0.00}
\definecolor{gray100}{rgb}{1.00,1.00,1.00}
\definecolor{gray10}{rgb}{0.10,0.10,0.10}
\definecolor{gray11}{rgb}{0.11,0.11,0.11}
\definecolor{gray12}{rgb}{0.12,0.12,0.12}
\definecolor{gray13}{rgb}{0.13,0.13,0.13}
\definecolor{gray14}{rgb}{0.14,0.14,0.14}
\definecolor{gray15}{rgb}{0.15,0.15,0.15}
\definecolor{gray16}{rgb}{0.16,0.16,0.16}
\definecolor{gray17}{rgb}{0.17,0.17,0.17}
\definecolor{gray18}{rgb}{0.18,0.18,0.18}
\definecolor{gray19}{rgb}{0.19,0.19,0.19}
\definecolor{gray1}{rgb}{0.01,0.01,0.01}
\definecolor{gray20}{rgb}{0.20,0.20,0.20}
\definecolor{gray21}{rgb}{0.21,0.21,0.21}
\definecolor{gray22}{rgb}{0.22,0.22,0.22}
\definecolor{gray23}{rgb}{0.23,0.23,0.23}
\definecolor{gray24}{rgb}{0.24,0.24,0.24}
\definecolor{gray25}{rgb}{0.25,0.25,0.25}
\definecolor{gray26}{rgb}{0.26,0.26,0.26}
\definecolor{gray27}{rgb}{0.27,0.27,0.27}
\definecolor{gray28}{rgb}{0.28,0.28,0.28}
\definecolor{gray29}{rgb}{0.29,0.29,0.29}
\definecolor{gray2}{rgb}{0.02,0.02,0.02}
\definecolor{gray30}{rgb}{0.30,0.30,0.30}
\definecolor{gray31}{rgb}{0.31,0.31,0.31}
\definecolor{gray32}{rgb}{0.32,0.32,0.32}
\definecolor{gray33}{rgb}{0.33,0.33,0.33}
\definecolor{gray34}{rgb}{0.34,0.34,0.34}
\definecolor{gray35}{rgb}{0.35,0.35,0.35}
\definecolor{gray36}{rgb}{0.36,0.36,0.36}
\definecolor{gray37}{rgb}{0.37,0.37,0.37}
\definecolor{gray38}{rgb}{0.38,0.38,0.38}
\definecolor{gray39}{rgb}{0.39,0.39,0.39}
\definecolor{gray3}{rgb}{0.03,0.03,0.03}
\definecolor{gray40}{rgb}{0.40,0.40,0.40}
\definecolor{gray41}{rgb}{0.41,0.41,0.41}
\definecolor{gray42}{rgb}{0.42,0.42,0.42}
\definecolor{gray43}{rgb}{0.43,0.43,0.43}
\definecolor{gray44}{rgb}{0.44,0.44,0.44}
\definecolor{gray45}{rgb}{0.45,0.45,0.45}
\definecolor{gray46}{rgb}{0.46,0.46,0.46}
\definecolor{gray47}{rgb}{0.47,0.47,0.47}
\definecolor{gray48}{rgb}{0.48,0.48,0.48}
\definecolor{gray49}{rgb}{0.49,0.49,0.49}
\definecolor{gray4}{rgb}{0.04,0.04,0.04}
\definecolor{gray50}{rgb}{0.50,0.50,0.50}
\definecolor{gray51}{rgb}{0.51,0.51,0.51}
\definecolor{gray52}{rgb}{0.52,0.52,0.52}
\definecolor{gray53}{rgb}{0.53,0.53,0.53}
\definecolor{gray54}{rgb}{0.54,0.54,0.54}
\definecolor{gray55}{rgb}{0.55,0.55,0.55}
\definecolor{gray56}{rgb}{0.56,0.56,0.56}
\definecolor{gray57}{rgb}{0.57,0.57,0.57}
\definecolor{gray58}{rgb}{0.58,0.58,0.58}
\definecolor{gray59}{rgb}{0.59,0.59,0.59}
\definecolor{gray5}{rgb}{0.05,0.05,0.05}
\definecolor{gray60}{rgb}{0.60,0.60,0.60}
\definecolor{gray61}{rgb}{0.61,0.61,0.61}
\definecolor{gray62}{rgb}{0.62,0.62,0.62}
\definecolor{gray63}{rgb}{0.63,0.63,0.63}
\definecolor{gray64}{rgb}{0.64,0.64,0.64}
\definecolor{gray65}{rgb}{0.65,0.65,0.65}
\definecolor{gray66}{rgb}{0.66,0.66,0.66}
\definecolor{gray67}{rgb}{0.67,0.67,0.67}
\definecolor{gray68}{rgb}{0.68,0.68,0.68}
\definecolor{gray69}{rgb}{0.69,0.69,0.69}
\definecolor{gray6}{rgb}{0.06,0.06,0.06}
\definecolor{gray70}{rgb}{0.70,0.70,0.70}
\definecolor{gray71}{rgb}{0.71,0.71,0.71}
\definecolor{gray72}{rgb}{0.72,0.72,0.72}
\definecolor{gray73}{rgb}{0.73,0.73,0.73}
\definecolor{gray74}{rgb}{0.74,0.74,0.74}
\definecolor{gray75}{rgb}{0.75,0.75,0.75}
\definecolor{gray76}{rgb}{0.76,0.76,0.76}
\definecolor{gray77}{rgb}{0.77,0.77,0.77}
\definecolor{gray78}{rgb}{0.78,0.78,0.78}
\definecolor{gray79}{rgb}{0.79,0.79,0.79}
\definecolor{gray7}{rgb}{0.07,0.07,0.07}
\definecolor{gray80}{rgb}{0.80,0.80,0.80}
\definecolor{gray81}{rgb}{0.81,0.81,0.81}
\definecolor{gray82}{rgb}{0.82,0.82,0.82}
\definecolor{gray83}{rgb}{0.83,0.83,0.83}
\definecolor{gray84}{rgb}{0.84,0.84,0.84}
\definecolor{gray85}{rgb}{0.85,0.85,0.85}
\definecolor{gray86}{rgb}{0.86,0.86,0.86}
\definecolor{gray87}{rgb}{0.87,0.87,0.87}
\definecolor{gray88}{rgb}{0.88,0.88,0.88}
\definecolor{gray89}{rgb}{0.89,0.89,0.89}
\definecolor{gray8}{rgb}{0.08,0.08,0.08}
\definecolor{gray90}{rgb}{0.90,0.90,0.90}
\definecolor{gray91}{rgb}{0.91,0.91,0.91}
\definecolor{gray92}{rgb}{0.92,0.92,0.92}
\definecolor{gray93}{rgb}{0.93,0.93,0.93}
\definecolor{gray94}{rgb}{0.94,0.94,0.94}
\definecolor{gray95}{rgb}{0.95,0.95,0.95}
\definecolor{gray96}{rgb}{0.96,0.96,0.96}
\definecolor{gray97}{rgb}{0.97,0.97,0.97}
\definecolor{gray98}{rgb}{0.98,0.98,0.98}
\definecolor{gray99}{rgb}{0.99,0.99,0.99}
\definecolor{gray9}{rgb}{0.09,0.09,0.09}
\definecolor{gray}{rgb}{0.75,0.75,0.75}
\definecolor{green1}{rgb}{0.00,1.00,0.00}
\definecolor{green2}{rgb}{0.00,0.93,0.00}
\definecolor{green3}{rgb}{0.00,0.80,0.00}
\definecolor{green4}{rgb}{0.00,0.55,0.00}
\definecolor{greenyellow}{rgb}{0.68,1.00,0.18}
\definecolor{green}{rgb}{0.00,1.00,0.00}
\definecolor{grey0}{rgb}{0.00,0.00,0.00}
\definecolor{grey100}{rgb}{1.00,1.00,1.00}
\definecolor{grey10}{rgb}{0.10,0.10,0.10}
\definecolor{grey11}{rgb}{0.11,0.11,0.11}
\definecolor{grey12}{rgb}{0.12,0.12,0.12}
\definecolor{grey13}{rgb}{0.13,0.13,0.13}
\definecolor{grey14}{rgb}{0.14,0.14,0.14}
\definecolor{grey15}{rgb}{0.15,0.15,0.15}
\definecolor{grey16}{rgb}{0.16,0.16,0.16}
\definecolor{grey17}{rgb}{0.17,0.17,0.17}
\definecolor{grey18}{rgb}{0.18,0.18,0.18}
\definecolor{grey19}{rgb}{0.19,0.19,0.19}
\definecolor{grey1}{rgb}{0.01,0.01,0.01}
\definecolor{grey20}{rgb}{0.20,0.20,0.20}
\definecolor{grey21}{rgb}{0.21,0.21,0.21}
\definecolor{grey22}{rgb}{0.22,0.22,0.22}
\definecolor{grey23}{rgb}{0.23,0.23,0.23}
\definecolor{grey24}{rgb}{0.24,0.24,0.24}
\definecolor{grey25}{rgb}{0.25,0.25,0.25}
\definecolor{grey26}{rgb}{0.26,0.26,0.26}
\definecolor{grey27}{rgb}{0.27,0.27,0.27}
\definecolor{grey28}{rgb}{0.28,0.28,0.28}
\definecolor{grey29}{rgb}{0.29,0.29,0.29}
\definecolor{grey2}{rgb}{0.02,0.02,0.02}
\definecolor{grey30}{rgb}{0.30,0.30,0.30}
\definecolor{grey31}{rgb}{0.31,0.31,0.31}
\definecolor{grey32}{rgb}{0.32,0.32,0.32}
\definecolor{grey33}{rgb}{0.33,0.33,0.33}
\definecolor{grey34}{rgb}{0.34,0.34,0.34}
\definecolor{grey35}{rgb}{0.35,0.35,0.35}
\definecolor{grey36}{rgb}{0.36,0.36,0.36}
\definecolor{grey37}{rgb}{0.37,0.37,0.37}
\definecolor{grey38}{rgb}{0.38,0.38,0.38}
\definecolor{grey39}{rgb}{0.39,0.39,0.39}
\definecolor{grey3}{rgb}{0.03,0.03,0.03}
\definecolor{grey40}{rgb}{0.40,0.40,0.40}
\definecolor{grey41}{rgb}{0.41,0.41,0.41}
\definecolor{grey42}{rgb}{0.42,0.42,0.42}
\definecolor{grey43}{rgb}{0.43,0.43,0.43}
\definecolor{grey44}{rgb}{0.44,0.44,0.44}
\definecolor{grey45}{rgb}{0.45,0.45,0.45}
\definecolor{grey46}{rgb}{0.46,0.46,0.46}
\definecolor{grey47}{rgb}{0.47,0.47,0.47}
\definecolor{grey48}{rgb}{0.48,0.48,0.48}
\definecolor{grey49}{rgb}{0.49,0.49,0.49}
\definecolor{grey4}{rgb}{0.04,0.04,0.04}
\definecolor{grey50}{rgb}{0.50,0.50,0.50}
\definecolor{grey51}{rgb}{0.51,0.51,0.51}
\definecolor{grey52}{rgb}{0.52,0.52,0.52}
\definecolor{grey53}{rgb}{0.53,0.53,0.53}
\definecolor{grey54}{rgb}{0.54,0.54,0.54}
\definecolor{grey55}{rgb}{0.55,0.55,0.55}
\definecolor{grey56}{rgb}{0.56,0.56,0.56}
\definecolor{grey57}{rgb}{0.57,0.57,0.57}
\definecolor{grey58}{rgb}{0.58,0.58,0.58}
\definecolor{grey59}{rgb}{0.59,0.59,0.59}
\definecolor{grey5}{rgb}{0.05,0.05,0.05}
\definecolor{grey60}{rgb}{0.60,0.60,0.60}
\definecolor{grey61}{rgb}{0.61,0.61,0.61}
\definecolor{grey62}{rgb}{0.62,0.62,0.62}
\definecolor{grey63}{rgb}{0.63,0.63,0.63}
\definecolor{grey64}{rgb}{0.64,0.64,0.64}
\definecolor{grey65}{rgb}{0.65,0.65,0.65}
\definecolor{grey66}{rgb}{0.66,0.66,0.66}
\definecolor{grey67}{rgb}{0.67,0.67,0.67}
\definecolor{grey68}{rgb}{0.68,0.68,0.68}
\definecolor{grey69}{rgb}{0.69,0.69,0.69}
\definecolor{grey6}{rgb}{0.06,0.06,0.06}
\definecolor{grey70}{rgb}{0.70,0.70,0.70}
\definecolor{grey71}{rgb}{0.71,0.71,0.71}
\definecolor{grey72}{rgb}{0.72,0.72,0.72}
\definecolor{grey73}{rgb}{0.73,0.73,0.73}
\definecolor{grey74}{rgb}{0.74,0.74,0.74}
\definecolor{grey75}{rgb}{0.75,0.75,0.75}
\definecolor{grey76}{rgb}{0.76,0.76,0.76}
\definecolor{grey77}{rgb}{0.77,0.77,0.77}
\definecolor{grey78}{rgb}{0.78,0.78,0.78}
\definecolor{grey79}{rgb}{0.79,0.79,0.79}
\definecolor{grey7}{rgb}{0.07,0.07,0.07}
\definecolor{grey80}{rgb}{0.80,0.80,0.80}
\definecolor{grey81}{rgb}{0.81,0.81,0.81}
\definecolor{grey82}{rgb}{0.82,0.82,0.82}
\definecolor{grey83}{rgb}{0.83,0.83,0.83}
\definecolor{grey84}{rgb}{0.84,0.84,0.84}
\definecolor{grey85}{rgb}{0.85,0.85,0.85}
\definecolor{grey86}{rgb}{0.86,0.86,0.86}
\definecolor{grey87}{rgb}{0.87,0.87,0.87}
\definecolor{grey88}{rgb}{0.88,0.88,0.88}
\definecolor{grey89}{rgb}{0.89,0.89,0.89}
\definecolor{grey8}{rgb}{0.08,0.08,0.08}
\definecolor{grey90}{rgb}{0.90,0.90,0.90}
\definecolor{grey91}{rgb}{0.91,0.91,0.91}
\definecolor{grey92}{rgb}{0.92,0.92,0.92}
\definecolor{grey93}{rgb}{0.93,0.93,0.93}
\definecolor{grey94}{rgb}{0.94,0.94,0.94}
\definecolor{grey95}{rgb}{0.95,0.95,0.95}
\definecolor{grey96}{rgb}{0.96,0.96,0.96}
\definecolor{grey97}{rgb}{0.97,0.97,0.97}
\definecolor{grey98}{rgb}{0.98,0.98,0.98}
\definecolor{grey99}{rgb}{0.99,0.99,0.99}
\definecolor{grey9}{rgb}{0.09,0.09,0.09}
\definecolor{grey}{rgb}{0.75,0.75,0.75}
\definecolor{honeydew1}{rgb}{0.94,1.00,0.94}
\definecolor{honeydew2}{rgb}{0.88,0.93,0.88}
\definecolor{honeydew3}{rgb}{0.76,0.80,0.76}
\definecolor{honeydew4}{rgb}{0.51,0.55,0.51}
\definecolor{honeydew}{rgb}{0.94,1.00,0.94}
\definecolor{hotpink}{rgb}{1.00,0.41,0.71}
\definecolor{indianred}{rgb}{0.80,0.36,0.36}
\definecolor{ivory1}{rgb}{1.00,1.00,0.94}
\definecolor{ivory2}{rgb}{0.93,0.93,0.88}
\definecolor{ivory3}{rgb}{0.80,0.80,0.76}
\definecolor{ivory4}{rgb}{0.55,0.55,0.51}
\definecolor{ivory}{rgb}{1.00,1.00,0.94}
\definecolor{khaki1}{rgb}{1.00,0.96,0.56}
\definecolor{khaki2}{rgb}{0.93,0.90,0.52}
\definecolor{khaki3}{rgb}{0.80,0.78,0.45}
\definecolor{khaki4}{rgb}{0.55,0.53,0.31}
\definecolor{khaki}{rgb}{0.94,0.90,0.55}
\definecolor{lavenderblush}{rgb}{1.00,0.94,0.96}
\definecolor{lavender}{rgb}{0.90,0.90,0.98}
\definecolor{lawngreen}{rgb}{0.49,0.99,0.00}
\definecolor{lemonchiffon}{rgb}{1.00,0.98,0.80}
\definecolor{lightblue}{rgb}{0.68,0.85,0.90}
\definecolor{lightcoral}{rgb}{0.94,0.50,0.50}
\definecolor{lightcyan}{rgb}{0.88,1.00,1.00}
\definecolor{lightgoldenrod}{rgb}{0.93,0.87,0.51}
\definecolor{lightgoldenrod}{rgb}{0.98,0.98,0.82}
\definecolor{lightgray}{rgb}{0.83,0.83,0.83}
\definecolor{lightgreen}{rgb}{0.56,0.93,0.56}
\definecolor{lightgrey}{rgb}{0.83,0.83,0.83}
\definecolor{lightpink}{rgb}{1.00,0.71,0.76}
\definecolor{lightsalmon}{rgb}{1.00,0.63,0.48}
\definecolor{lightsea}{rgb}{0.13,0.70,0.67}
\definecolor{lightsky}{rgb}{0.53,0.81,0.98}
\definecolor{lightslate}{rgb}{0.47,0.53,0.60}
\definecolor{lightslate}{rgb}{0.47,0.53,0.60}
\definecolor{lightslate}{rgb}{0.52,0.44,1.00}
\definecolor{lightsteel}{rgb}{0.69,0.77,0.87}
\definecolor{lightyellow}{rgb}{1.00,1.00,0.88}
\definecolor{limegreen}{rgb}{0.20,0.80,0.20}
\definecolor{linen}{rgb}{0.98,0.94,0.90}
\definecolor{magenta1}{rgb}{1.00,0.00,1.00}
\definecolor{magenta2}{rgb}{0.93,0.00,0.93}
\definecolor{magenta3}{rgb}{0.80,0.00,0.80}
\definecolor{magenta4}{rgb}{0.55,0.00,0.55}
\definecolor{magenta}{rgb}{1.00,0.00,1.00}
\definecolor{maroon1}{rgb}{1.00,0.20,0.70}
\definecolor{maroon2}{rgb}{0.93,0.19,0.65}
\definecolor{maroon3}{rgb}{0.80,0.16,0.56}
\definecolor{maroon4}{rgb}{0.55,0.11,0.38}
\definecolor{maroon}{rgb}{0.69,0.19,0.38}
\definecolor{mediumaquamarine}{rgb}{0.40,0.80,0.67}
\definecolor{mediumblue}{rgb}{0.00,0.00,0.80}
\definecolor{mediumorchid}{rgb}{0.73,0.33,0.83}
\definecolor{mediumpurple}{rgb}{0.58,0.44,0.86}
\definecolor{mediumsea}{rgb}{0.24,0.70,0.44}
\definecolor{mediumslate}{rgb}{0.48,0.41,0.93}
\definecolor{mediumspring}{rgb}{0.00,0.98,0.60}
\definecolor{mediumturquoise}{rgb}{0.28,0.82,0.80}
\definecolor{mediumviolet}{rgb}{0.78,0.08,0.52}
\definecolor{midnightblue}{rgb}{0.10,0.10,0.44}
\definecolor{mintcream}{rgb}{0.96,1.00,0.98}
\definecolor{mistyrose}{rgb}{1.00,0.89,0.88}
\definecolor{moccasin}{rgb}{1.00,0.89,0.71}
\definecolor{navajowhite}{rgb}{1.00,0.87,0.68}
\definecolor{navyblue}{rgb}{0.00,0.00,0.50}
\definecolor{navy}{rgb}{0.00,0.00,0.50}
\definecolor{oldlace}{rgb}{0.99,0.96,0.90}
\definecolor{olivedrab}{rgb}{0.42,0.56,0.14}
\definecolor{orange1}{rgb}{1.00,0.65,0.00}
\definecolor{orange2}{rgb}{0.93,0.60,0.00}
\definecolor{orange3}{rgb}{0.80,0.52,0.00}
\definecolor{orange4}{rgb}{0.55,0.35,0.00}
\definecolor{orangered}{rgb}{1.00,0.27,0.00}
\definecolor{orange}{rgb}{1.00,0.65,0.00}
\definecolor{orchid1}{rgb}{1.00,0.51,0.98}
\definecolor{orchid2}{rgb}{0.93,0.48,0.91}
\definecolor{orchid3}{rgb}{0.80,0.41,0.79}
\definecolor{orchid4}{rgb}{0.55,0.28,0.54}
\definecolor{orchid}{rgb}{0.85,0.44,0.84}
\definecolor{palegoldenrod}{rgb}{0.93,0.91,0.67}
\definecolor{palegreen}{rgb}{0.60,0.98,0.60}
\definecolor{paleturquoise}{rgb}{0.69,0.93,0.93}
\definecolor{paleviolet}{rgb}{0.86,0.44,0.58}
\definecolor{papayawhip}{rgb}{1.00,0.94,0.84}
\definecolor{peachpuff}{rgb}{1.00,0.85,0.73}
\definecolor{peru}{rgb}{0.80,0.52,0.25}
\definecolor{pink1}{rgb}{1.00,0.71,0.77}
\definecolor{pink2}{rgb}{0.93,0.66,0.72}
\definecolor{pink3}{rgb}{0.80,0.57,0.62}
\definecolor{pink4}{rgb}{0.55,0.39,0.42}
\definecolor{pink}{rgb}{1.00,0.75,0.80}
\definecolor{plum1}{rgb}{1.00,0.73,1.00}
\definecolor{plum2}{rgb}{0.93,0.68,0.93}
\definecolor{plum3}{rgb}{0.80,0.59,0.80}
\definecolor{plum4}{rgb}{0.55,0.40,0.55}
\definecolor{plum}{rgb}{0.87,0.63,0.87}
\definecolor{powderblue}{rgb}{0.69,0.88,0.90}
\definecolor{purple1}{rgb}{0.61,0.19,1.00}
\definecolor{purple2}{rgb}{0.57,0.17,0.93}
\definecolor{purple3}{rgb}{0.49,0.15,0.80}
\definecolor{purple4}{rgb}{0.33,0.10,0.55}
\definecolor{purple}{rgb}{0.63,0.13,0.94}
\definecolor{red1}{rgb}{1.00,0.00,0.00}
\definecolor{red2}{rgb}{0.93,0.00,0.00}
\definecolor{red3}{rgb}{0.80,0.00,0.00}
\definecolor{red4}{rgb}{0.55,0.00,0.00}
\definecolor{red}{rgb}{1.00,0.00,0.00}
\definecolor{rosybrown}{rgb}{0.74,0.56,0.56}
\definecolor{royalblue}{rgb}{0.25,0.41,0.88}
\definecolor{saddlebrown}{rgb}{0.55,0.27,0.07}
\definecolor{salmon1}{rgb}{1.00,0.55,0.41}
\definecolor{salmon2}{rgb}{0.93,0.51,0.38}
\definecolor{salmon3}{rgb}{0.80,0.44,0.33}
\definecolor{salmon4}{rgb}{0.55,0.30,0.22}
\definecolor{salmon}{rgb}{0.98,0.50,0.45}
\definecolor{sandybrown}{rgb}{0.96,0.64,0.38}
\definecolor{seagreen}{rgb}{0.18,0.55,0.34}
\definecolor{seashell1}{rgb}{1.00,0.96,0.93}
\definecolor{seashell2}{rgb}{0.93,0.90,0.87}
\definecolor{seashell3}{rgb}{0.80,0.77,0.75}
\definecolor{seashell4}{rgb}{0.55,0.53,0.51}
\definecolor{seashell}{rgb}{1.00,0.96,0.93}
\definecolor{sienna1}{rgb}{1.00,0.51,0.28}
\definecolor{sienna2}{rgb}{0.93,0.47,0.26}
\definecolor{sienna3}{rgb}{0.80,0.41,0.22}
\definecolor{sienna4}{rgb}{0.55,0.28,0.15}
\definecolor{sienna}{rgb}{0.63,0.32,0.18}
\definecolor{skyblue}{rgb}{0.53,0.81,0.92}
\definecolor{slateblue}{rgb}{0.42,0.35,0.80}
\definecolor{slategray}{rgb}{0.44,0.50,0.56}
\definecolor{slategrey}{rgb}{0.44,0.50,0.56}
\definecolor{snow1}{rgb}{1.00,0.98,0.98}
\definecolor{snow2}{rgb}{0.93,0.91,0.91}
\definecolor{snow3}{rgb}{0.80,0.79,0.79}
\definecolor{snow4}{rgb}{0.55,0.54,0.54}
\definecolor{snow}{rgb}{1.00,0.98,0.98}
\definecolor{springgreen}{rgb}{0.00,1.00,0.50}
\definecolor{steelblue}{rgb}{0.27,0.51,0.71}
\definecolor{tan1}{rgb}{1.00,0.65,0.31}
\definecolor{tan2}{rgb}{0.93,0.60,0.29}
\definecolor{tan3}{rgb}{0.80,0.52,0.25}
\definecolor{tan4}{rgb}{0.55,0.35,0.17}
\definecolor{tan}{rgb}{0.82,0.71,0.55}
\definecolor{thistle1}{rgb}{1.00,0.88,1.00}
\definecolor{thistle2}{rgb}{0.93,0.82,0.93}
\definecolor{thistle3}{rgb}{0.80,0.71,0.80}
\definecolor{thistle4}{rgb}{0.55,0.48,0.55}
\definecolor{thistle}{rgb}{0.85,0.75,0.85}
\definecolor{tomato1}{rgb}{1.00,0.39,0.28}
\definecolor{tomato2}{rgb}{0.93,0.36,0.26}
\definecolor{tomato3}{rgb}{0.80,0.31,0.22}
\definecolor{tomato4}{rgb}{0.55,0.21,0.15}
\definecolor{tomato}{rgb}{1.00,0.39,0.28}
\definecolor{turquoise1}{rgb}{0.00,0.96,1.00}
\definecolor{turquoise2}{rgb}{0.00,0.90,0.93}
\definecolor{turquoise3}{rgb}{0.00,0.77,0.80}
\definecolor{turquoise4}{rgb}{0.00,0.53,0.55}
\definecolor{turquoise}{rgb}{0.25,0.88,0.82}
\definecolor{violetred}{rgb}{0.82,0.13,0.56}
\definecolor{violet}{rgb}{0.93,0.51,0.93}
\definecolor{wheat1}{rgb}{1.00,0.91,0.73}
\definecolor{wheat2}{rgb}{0.93,0.85,0.68}
\definecolor{wheat3}{rgb}{0.80,0.73,0.59}
\definecolor{wheat4}{rgb}{0.55,0.49,0.40}
\definecolor{wheat}{rgb}{0.96,0.87,0.70}
\definecolor{whitesmoke}{rgb}{0.96,0.96,0.96}
\definecolor{white}{rgb}{1.00,1.00,1.00}
\definecolor{yellow1}{rgb}{1.00,1.00,0.00}
\definecolor{yellow2}{rgb}{0.93,0.93,0.00}
\definecolor{yellow3}{rgb}{0.80,0.80,0.00}
\definecolor{yellow4}{rgb}{0.55,0.55,0.00}
\definecolor{yellowgreen}{rgb}{0.60,0.80,0.20}
\definecolor{yellow}{rgb}{1.00,1.00,0.00}

\newcolumntype{n}{>{\raggedleft \arraybackslash} p{0.9cm}}
\newcolumntype{m}{>{\raggedleft \arraybackslash} p{1.3cm}}

\begin{document}
\mainmatter
  
\title{How Advanced Change Patterns Impact the Process of Process
Modeling\thanks{This research is supported by Austrian Science Fund (FWF):
P23699-N23}}
\titlerunning{How Advanced Change Patterns Impact the Process of Process
Modeling} 
\author{Barbara Weber\inst{1} \and Sarah Zeitlhofer\inst{1} \and Jakob
Pinggera\inst{1} \and Victoria Torres\inst{2} \and Manfred Reichert\inst{3}}
\authorrunning{Weber et al.}
 
\institute{University of Innsbruck, Austria\\
\email{firstname.lastname@uibk.ac.at |
sarah.zeitlhofer@student.uibk.ac.at} \and Universitat Polit\`ecnica de
Val\`encia, Spain\\
\email{vtorres@pros.upv.es}
\and University of Ulm, Germany\\
\email{manfred.reichert@uni-ulm.de} 
} 

\maketitle

\begin{abstract}
Process model quality has been an area of considerable research efforts. In this
context, correctness-by-construction as enabled by change patterns provides promising
perspectives. While the process of process modeling (PPM) based on change
primitives has been thoroughly investigated, only little is known about the PPM based on
change patterns. In particular, it is unclear what set of change patterns should
be provided and how the available change pattern set impacts the PPM. To obtain
a better understanding of the latter as well as the (subjective) perceptions of
process modelers, the arising challenges, and the pros and cons of different
change pattern sets we conduct a controlled experiment. Our results indicate
that process modelers face similar challenges irrespective of the used change
pattern set (core pattern set versus extended pattern set, which 
adds two advanced change patterns to the core patterns set). An extended change
pattern set, however, is perceived as more difficult to use, yielding a higher
mental  effort. Moreover, our results indicate that more advanced patterns were 
only used to a limited extent and frequently applied incorrectly, thus, lowering
the potential benefits of an extended pattern set.
\end{abstract}

\section{Introduction} 
Due to the important role they play for process--aware information
 systems, process models have become increasingly important for many years~\cite{Bec+00}. In this context, it was shown that process model
 understandability has a measurable impact on whether or not a process modeling initiative is
 successful~\cite{DBLP:journals/dss/KockVDD09}. Still, process models exhibit a
 wide range of quality problems, which not only hamper comprehensibility
 but also affect maintainability~\cite{DBLP:journals/dke/MendlingVDAN08,WRR08}.
 For example, \cite{DBLP:journals/dke/MendlingVDAN08} reports on error rates
 between 10\% and 20\% in collections of industrial process models.

To improve process model quality, change patterns appear promising. They combine change primitives, e.g., to add nodes or edges, to
high-level change operations~\cite{WRR08}. In particular,
change patterns enable correctness-by-construction~\cite{Casa98} by providing
only those change patterns to the modeler, which ensure that process
models remain sound after applying model transformations.

Recently, the creation of process models based on change primitives has
received considerable attention resulting in research on the
\textit{process of process modeling (PPM)}~\cite{PZW+12,CVR+12,PSZ+13}. This research focuses on the \emph{formalization} phase of process
model creation, i.e., the interactions of the process modeler with the modeling
environment. The PPM utilizing change patterns, in turn, is still hardly
understood. In previous work we presented an exploratory study to investigate
re-occurring challenges when using change patterns for process
modeling~\cite{WPVR13}. The study revealed that process modelers did not face
major problems when using change patterns for constructing simple process
fragments. When being confronted with more complex process fragments, however,
difficulties increased observably. Building respective structures efficiently
(i.e., without \emph{detours} in the PPM) requires process modelers to look
ahead, since patterns cannot be always combined arbitrarily. This need for
looking ahead is a fundamental difference compared to process model
creation based on change primitives and was perceived as both challenging and
restrictive by subjects. Further, \cite{WPVR13} emphasizes that the basic set of
change patterns, which allows creating control flow structures like
sequence, exclusive branchings, parallel branchings, and loops, is not
sufficient for efficient model creation. In particular, the study observed that
patterns for moving process fragments might help to resolve detours
efficiently.

On one hand, an extended set of change patterns (including move patterns)
offers more flexibility. On the other, it  increases complexity, especially when
mapping the mental model of the process to be created to the available pattern
set. As a result, the extended change pattern set might make the modeling
environment more difficult to use. This raises the question whether the expected
benefits of an extended pattern set can be materialized in a practical setting.
To obtain an in-depth understanding of the impact an extended pattern set has on
the PPM, we implement a modeling editor offering two different change pattern
sets based on Cheetah Experimental Platform (CEP)~\cite{PZW10}.
Using this editor, we conduct a controlled experiment with 42 process modelers.
Our results indicate that an extended pattern set yields higher mental effort
for modelers and is perceived as more difficult to use. At the same time, the
expected benefits in terms of increased problem solving efficiency did not
materialize, suggesting to focus on a core pattern set. The results provide a
contribution toward a better understanding on how tool features (like change
patterns) impact the PPM, but also give advice on how effective tool support
should be designed.

Sect.~\ref{backgrounds} introduces backgrounds. Sect.~\ref{exploratoryStudy}
describes the controlled experiment. Sect.~\ref{easeOfUse} presents the
subjective perception of change pattern use. Sect.~\ref{challenges} deals with
the impact of change patterns on problem solving efficiency and
Sect.~\ref{advantagesAndDisadvantages} details on the actual and potential
use of patterns. Limitations are presented in Sect.~\ref{discussion}.
Related work is presented in Sect.~\ref{relatedWork}. Sect.~\ref{summary}
concludes the paper.

\vspace{-0.5cm}

\section{Process Model Creation Based on Change Patterns}
\label{backgrounds}

Most environments for process model creation are based on
change primitives, e.g., {\tt add/delete activity} or {\tt add/delete edge}.
Process model adaptations (i.e., transformation of a model S into model S') may require the joint application of multiple change primitives. Imagine process model $S_1$ in Fig.~\ref{img:example} without the colored fragment. To transform this model into S1 (including the colored fragment) 19 change primitives are needed: deleting the edge
between activity \texttt{D} and the parallel gateway, adding \texttt{D},\texttt{E}, and \texttt{F} to the process model, adding the conditional branch around \texttt{C} (including transition conditions), and adding the edges connecting the newly added
elements with the process model. When applying change primitive, soundness
of the resulting process model cannot be guaranteed and must be explicitly
checked after every model transformation.
In turn, change patterns imply a different
way of interacting with the modeling environment. Instead of applying a set of
change primitives, high-level change operations are used to realize the desired
model transformation. Examples of change patterns include the insertion of
process fragments, their embedding in conditional branches or loops, or the
updating of transition conditions. A catalog of change patterns can be found in
\cite{WRR08}, while their semantics of these patterns are described
in~\cite{DBLP:conf/er/Rinderle-MaRW08}. To conduct the described transformation with change patterns (i.e., obtain $S_1$ from a model where the colored fragment is missing), 6 pattern applications are needed (i.e., serial insert of activity \texttt{E}, parallel insert of
activity \texttt{F},  serial insert of activity \texttt{C}, embed activity
\texttt{C} in conditional branch, and two updates of conditions).
As opposed to change primitives, change pattern implementations typically
guarantee model correctness after each transformation \cite{Casa98} by
associating pre-/post-conditions with high-level change operations. In process
modeling environments supporting the correctness--by--construction principle
(e.g., \cite{DaRe09}), usually process modelers only have those change patterns
available for use that allow transforming a sound process model into another
sound one. For this purpose, structural restrictions on
process models (e.g., block structuredness) are imposed.  This
paper investigates the impact of two different change pattern sets on the PPM.

\vspace{-0.4cm}
\section{Experiment}
\label{exploratoryStudy}
This section describes research questions and the design of the experiment.

\textbf{Research Questions.} Our goal is to obtain an in-depth PPM understanding
when using change patterns. More specifically, we want to understand how
modelers experience their interaction with the modeling environment depending on
the available change pattern set.

\begin{center}
\fbox{
  \parbox{11cm}{
 RQ1: What is the impact of the change pattern set available to process modelers
on their subjective perception during model creation? }
 } 
\end{center} 

\noindent In addition to the subjective perception of modelers, we are
interested in the challenges faced by process modelers during the
PPM depending on the used change pattern set. Respective challenges can result
in modeling errors that persist in the final model, but also detours on the way
to a complete process model, negatively affecting problem solving efficiency.

\vspace{0.1cm}
\begin{center}
\fbox{
  \parbox{11cm}{
 RQ2: What is the impact of the change pattern set available to process modelers
 on the challenges faced during model creation? } } 
\end{center}
\vspace{0.1cm}

\noindent Finally, we want to understand how well the additional patterns of the extended pattern set was adopted (i.e., in their actual use) as well as the
potential benefits that could have been achieved through proper pattern usage.

\vspace{0.1cm}
\begin{center}
\fbox{
  \parbox{11cm}{
 RQ3: What was the actual use of the additional change patterns compared to the potential of using those patterns?
  }
 } 
\end{center}
\vspace{0.1cm}

\begin{figure}[h]
\begin{center}
\includegraphics[width=\textwidth]{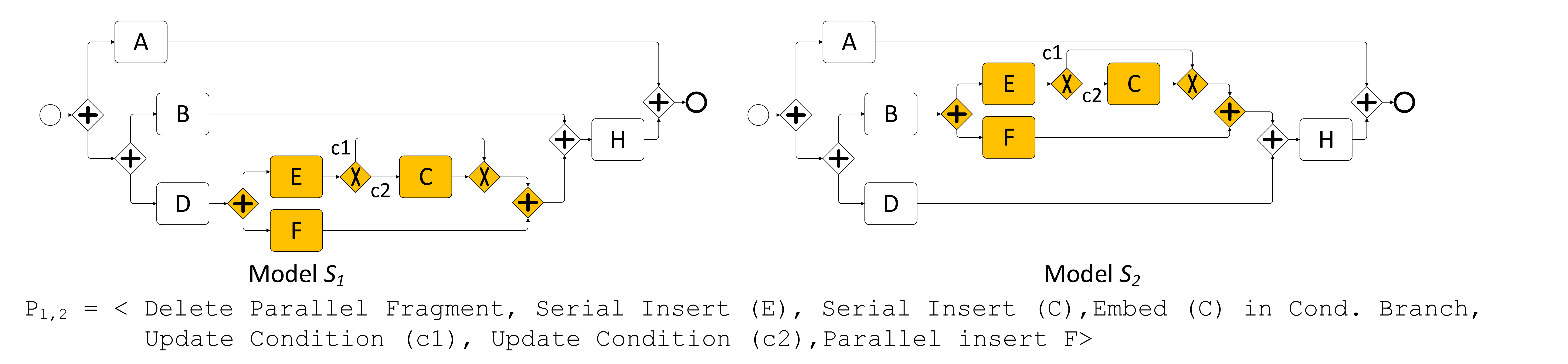}
\caption{Detour to go from $S_1$ to $S_2$ when no Move Patterns are available}
\label{img:example}
\end{center}
\vspace{-0.7cm}
\end{figure}

\textbf{Factors and Factor Level.} The experiment considers a single factor,
i.e, the pattern set used to conduct the modeling task with factor levels:
\textit{core} and \textit{extended}. The core pattern set comprises a minimum
change pattern set (see~\cite{WRR08} for the full pattern set) that allows
modelers to create basic control-flow structures (i.e., sequences, parallel,
conditional branchings, and loops): patterns AP1 (Insert Process
Fragment), AP2 (Delete Process Fragment), AP8 (Embed Fragment in Loop), AP10
(Embed Process Fragment in Conditional Branch), and AP13 (Update Condition).
Concerning pattern AP1, two variants were provided: Serial and Parallel Insert.
In addition, process modelers could rename activities.
In turn, the extended pattern set comprises all patterns included in the core
pattern set plus an advanced pattern for moving process fragments
(AP3). To be able to trace back the impact to single change patterns, we
intentionally decided to only add one additional pattern from which we expect
a considerable impact on problem solving efficiency to the extended pattern set.
Similar to AP1, two variants are provided: Serial and Parallel Move.
While the core pattern set is complete in the sense that all control-flow
structures can be created, it does not allow for arbitrary model
transformations. In particular, in~\cite{WPVR13} we observed that detours could
have been addressed more efficiently with an extended pattern set. In
particular, we observed that patterns for moving process fragments would have
helped with many of the detours. Frequently, process modelers had to undo or
delete considerable parts of the model, which could have been resolved with the
application of a single move pattern. Consider, for example, the two models in
Fig.~\ref{img:example}. When transforming $S_1$ to $S_2$ without move patterns, the
modeler must perform a detour of 7 steps to delete the colored parallel branch
and to re-insert it after activity B (cf. problem solving path $P_{1,2}$). On the contrary, using move patterns,
transforming $S_1$ into $S_2$ just requires the application of one change pattern,
i.e., Serial Move, saving a total of 6 pattern applications.

\begin{figure}[h]
\begin{center}
\includegraphics[width=\textwidth]{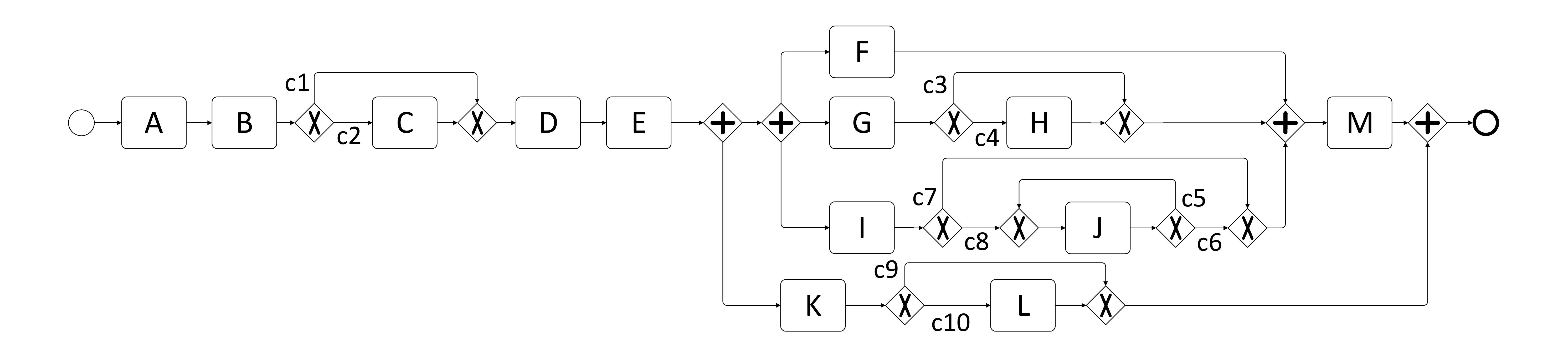}
\caption{Solution Model $S_S$}
\label{img:tasksA+B}
\end{center}
\end{figure}

\textbf{Modeling Tasks.} The modeling task is a slight adaption of the task used
in \cite{WPVR13} and describes a process of the ``Task Force
Earthquakes'' of the German Research Center for
Geosciences~\cite{DBLP:conf/bpm/FahlandW08} (cf.
Fig.~\ref{img:tasksA+B}---labels are abstracted for readability). The task
comprises 15 activities; all main control--flow structures like sequence,
parallel and conditional branchings, and loops are present. The model has a
nesting depth of 4. Subjects received an informal requirements description as
well as the solution of the modeling task (i.e., a process model). 
Their task was to re--model the process using change patterns. To
model the process a minimum number of 28 change patterns are required with both
the core and the extended change pattern set. Since subjects had the correct
solution available, the challenge lies in determining the patterns for
re-constructing the model and in combining the available patterns effectively.

\vspace{-0.5cm}
\vspace{-0.7cm}

\textbf{Subjects.} Novices and experts differ in their problem solving
strategies. Considering that industrial process modelers are often not expert modelers, but rather
casual modelers with a basic amount of training~\cite{PZW+10}, subjects participating 
in our experiment are not required to be experts. In previous research with software engineering students it has been shown that students can provide an adequate model for the professional
population~\cite{DBLP:journals/ese/HostRW00,DBLP:journals/ese/PorterV98,Run03}. Thus, we relied on students (instead of professionals) in our experiment. To avoid difficulties due to unfamiliarity with the tool, rather than the modeling task, we require some prior experience with process modeling as well as
change patterns. To ensure that the subjects are sufficiently literate in change
pattern usage, subjects are provided with theoretical backgrounds. Further the
subjects obtain hands-on experience in the creation of process models using
change patterns in terms of a familiarization task.

\textbf{Experimental Setup.} The experiment consists of four phases.
(1) collecting demographic data, (2) familiarization with the change patterns
editor, and (3) performing a modeling task. Subjects were divided into two
groups. While \emph{Group A} receives  the core pattern set, \emph{Group B}
conducts the same task based on the extended pattern set. During modeling, all
interactions with the modeling environment are recorded by CEP~\cite{PZW10}.
This allows us to replay the creation of the process model step-by-step
\cite{PZW10}, addressing RQ2 and RQ3. After completing the modeling tasks, (4)
\emph{mental effort} as well as \textit{Perceived Ease of Use (PEU)} and
\textit{Perceived Usefulness (PU)} of the \textit{Technology Acceptance
Model}~\cite{Davis89} are assessed, addressing RQ1.

\textbf{Experimental Execution.}  Prior to the experiment a pilot was conducted
to ensure usability of the tool and understandability of the task description.
This led to improvements of CEP and minor updates of the task description. The
experiment was conducted by 42 graduate and postgraduate students from the
Universities of Innsbruck, Ulm, and Valencia. Subjects were randomly assigned to
groups, with an equal number of subjects for each group.
 
\textbf{Data Validation.} To obtain a valid data set, we checked for
completeness of the created process models. Unfortunately, 8 of the participants
had to be removed due to incomplete models. As, a result, 34 subjects remained in
the data set, which were equally distributed over the two groups. Since we did not consider process modeling knowledge and experience as a factor
in our experiment, we screened the participants for prior knowledge regarding
BPMN and change patterns. For this, a questionnaire similar to~\cite{Men+07} was
used to verify that subjects were equally distributed to the two groups. (cf.
Table~\ref{tab:demographics}). The questionnaire used Likert scale ranging from
strongly disagree (1) to strongly agree (7). To test for differences between the
two groups, t-tests were run for normally distributed data. For non--normally
distributed data, the Mann-Whitney test was used. No significant differences
were identified between the two groups. Consequently, we conclude that no
differences could be observed between the two groups.

\begin{table}
\vspace{-0.3cm}
\centering
\begin{tabular}{lrnnnn} 
Question & Group & Min & Max & M & SD \\
\hline
Familiarity with BPMN & A & 2 & 7 & 5.12 & 0.99 \\
& B & 2 & 7 & 5.53 & 1.28\\
Confidence in understanding BPMN & A & 3 & 7 & 5.53 & 1.33\\
& B & 4 & 7 & 6.24 & 0.75\\
Competence using BPMN & A & 3 & 7 & 5.06 & 1.14\\
& B &  3 & 7 & 5.59 & 1.06 \\
Familiarity change patterns & A & 2 & 7 & 4.76 & 1.44\\
 & B & 2 & 7 & 4.53 & 1.46 \\
Competence using change patterns & A & 2 & 7 & 4.59 & 1.33\\
 & B & 2 & 6 & 4.41 & 1.28 \\
\end{tabular}
\caption{Demographic Data}
\label{tab:demographics}
\vspace{-1.2cm}
\end{table}

\section{Subjective Perception of Model Creation}
\label{easeOfUse}
This section addresses research question RQ1 dealing with the subjective
perception of process modelers when using change patterns. In particular, it
investigates how the used change pattern set (core vs.
extended) impacts mental effort. Further, we investigate the perceived ease of
use and perceived usefulness.
 
\vspace{-0.2cm}
\subsection{Mental Effort} 
\label{mentalEffort} 

\paragraph{Descriptive Statistics.}
The results related to mental efforts are displayed in Table
\ref{tab:perceivedEaseOfUseAndPerceivedUsefulness}. Mental effort was measured
using a 7-point Likert scale ranging from 'very low' (1) to 'very high' (7). For
Group A the mean mental effort was 3.35, corresponding approx. to 'rather low'
(3). In turn, for Group B the mental effort was higher with a mean of
4.25, corresponding to 'medium' (4).

\begin{table}
\vspace{-0.5cm}
\centering
\begin{tabular}{lrnnnn}
Scale & Group & Min & Max & M & SD \\
\hline
Mental Effort & A & 2 & 5 & 3.35 & 1.06 \\
 & B & 3 & 7 & 4.25 & 1.00 \\
Perceived Ease of Use & A & 5.18 & 6.06 & 5.81 & 0.33 \\
 &  B & 4.53 & 5.82 & 5.25 & 0.47 \\
Perceived Usefulness &  A & 4.13 & 4.75 & 4.38 & 0.21 \\
 &  B & 3.87 & 4.87 & 4.27 & 0.36 \\
\end{tabular}
\caption{Subjective Perception}
\label{tab:perceivedEaseOfUseAndPerceivedUsefulness}
\vspace{-1cm}
\end{table}

\paragraph{Hypothesis Testing.}
When using change patterns for process modeling, plan schemata on how to apply
change patterns need to be developed in order to create complex process
fragments. In this context, we investigate how the mental effort of modelers is
affected by utilizing a larger change pattern set. While the extended change
pattern set allows modelers to recover faster from detours, it also requires
them to develop additional plan schemata on how to apply the move change
patterns. Therefore, an extended change pattern set might impose higher demands
on the modeler's cognitive resources. Especially, move change patterns require
modelers to imagine how the process model looks like after applying 
change patterns. This might put additional burden on them, requiring to
manipulate an internal representation of the process model in working memory. In
the light of the cognitive background, we expect the extended pattern set to
yield a significantly higher mental effort compared to the core pattern set.

\textbf{Hypothesis H1} \textit{The usage of an extended change pattern set significantly increases the mental effort required to accomplish the modeling task.}

Since the data was normally distributed, a t-test was used for testing the differences between the two groups ($t(31)=-2.50, p=0.02$). The result allows us
to accept hypothesis H1.

\label{tam}
\paragraph{Descriptive Statistics.}
In order to assess how far process modelers with moderate process modeling
knowledge consider the change pattern editor as easy to use and useful, we asked them
to fill out the \textit{Perceived Ease of Use (PEU)} and the \textit{Perceived
Usefulness (PU)}. Both scales consist of 7-point Likert items, ranging
from 'extremely unlikely' (1) over 'neither likely nor unlikely' (4) to
'extremely likely' (7). Regarding the PEU, the mean value was 5.81 for
Group A (core pattern set), corresponding approx. to 'quite
likely' (6). In turn, for Group B (extended pattern set) the mean value was
5.25, corresponding approx. to 'slightly likely' (5).
Finally, regarding the PU, the observed mean value was 4.38 for Group A and 4.27
for Group B, corresponding approx. to 'neither likely nor unlikely' (4) for both
groups. Three participants indicated that they could not answer the questions on
PU. Hence, they were removed for the analysis of PU. 

\paragraph{Hypothesis Testing.}
% As stated in the context of mental effort, the extended pattern set
% requires from modelers to develop additional plan schemata in order to apply the
% change patterns properly. Accordingly, one would expect that an extended change
% pattern set is even more difficult to use. On the contrary, process modelers
% might perceive the limited core pattern set as being restrictive, since
% several steps are required to resolve a detour. Particularly, detours might be
% resolved quicker when using the extended change pattern set, i.e., when
% allowing to move a misplaced process fragment based on a respective pattern.
% Therefore, one would expect that the extended change pattern set is
% perceived to be more useful.

As stated for mental effort already, the extended pattern set
requires modelers to develop additional plan schemata in order to apply the
change patterns properly. Accordingly, one would expect that an extended change
pattern set is more difficult to use. However, these should also be perceived
as more useful since the extended pattern set helps to resolve detours quicker
compared to the core pattern set, i.e., when allowing to move a misplaced process
fragment based on a respective pattern.

\textbf{Hypothesis H2} \textit{The usage of an extended change pattern set
significantly lowers the perceived ease of use.}

\textbf{Hypothesis H3} \textit{The usage of an extended change pattern set
significantly increases the perceived usefulness.}

Since none of the groups are normally distributed, we apply the Mann-Whitney
U-Test to test for differences regarding PEU and PU. While significant
differences in terms of PEU ($U=4010.50, p = 0.00$) allow us to accept
hypothesis H2, no statistically significant differences in terms of PU
($U=3639.00, p = 0.06$) were observed.

\paragraph{Discussion.}
Our results indicate that the core pattern set leads to a significantly lower
mental effort for modelers and its use is perceived as being
significantly easier compared to the extended pattern set.
This seems reasonable since modelers need to devote additional cognitive
resources in order to use the move change patterns. Regarding PU, against our
expectations, we could not obtain any statistically significant result. When
looking at the descriptive statistics, the participants of Group B tend to
perceive change patterns as less useful compared to Group A. We might conclude
that the move change patterns provided for Group B are not as useful as expected
(at least for the task assigned to the subjects).
Alternatively, the subjects of Group B might have struggled with the usage of
change patterns due to the additional patterns. In turn, this might have foiled
potential positive effects of the additional patterns. The results presented
in Sec.~\ref{challenges} support the latter explanation suggesting that process
modelers had considerable problems with the use of the move patterns.

\section{Challenges when Modeling with Change Patterns}
\label{challenges}
This section addresses research question RQ2 aiming to obtain an in-depth
understanding how the chosen pattern set impacts the challenges faced
by modelers.

\subsection{Data Analysis Procedure}  \label{dataAnalysis}
 
\begin{figure}[h]
\begin{center}
\includegraphics[width=\textwidth]{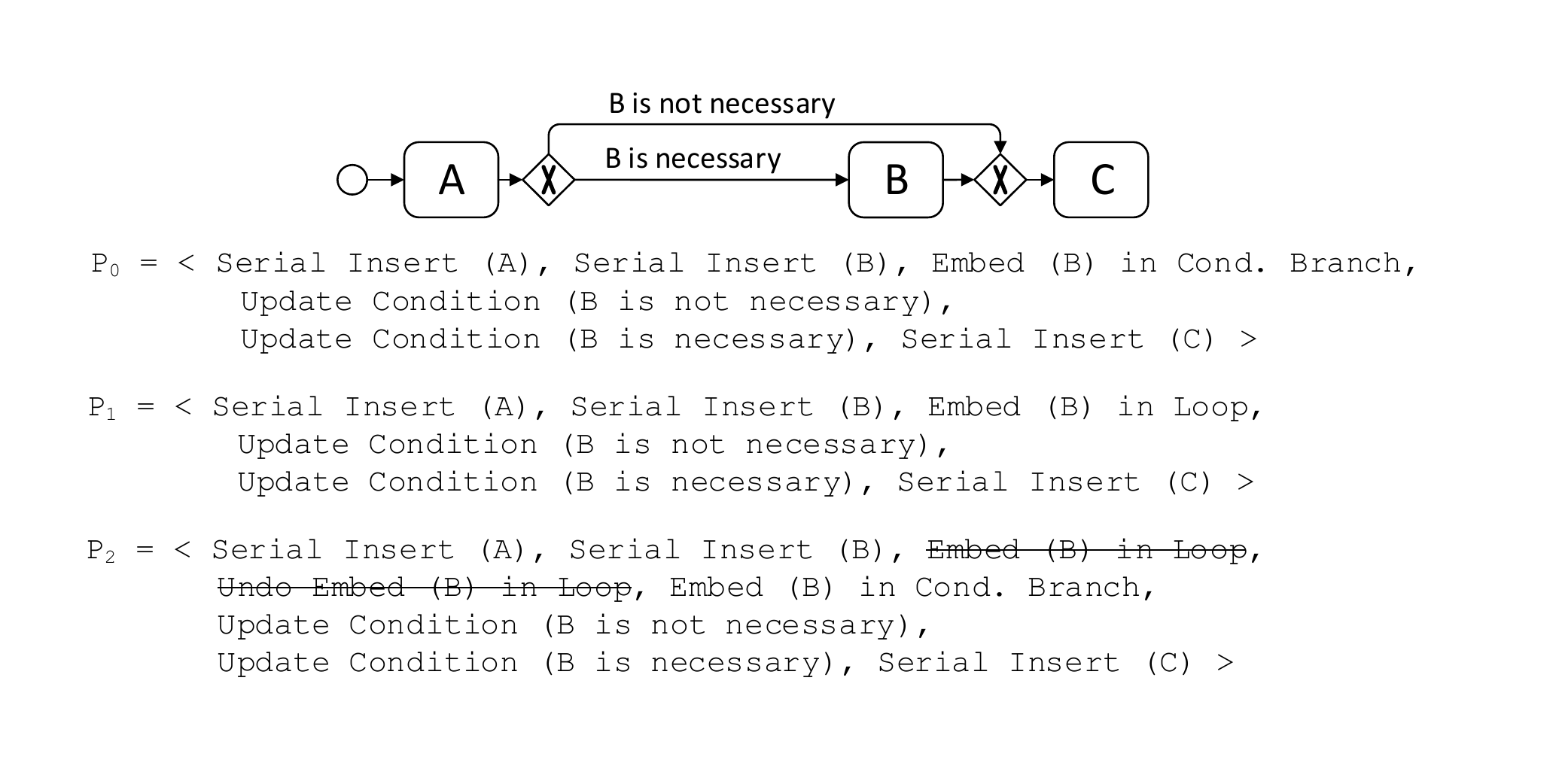}
\vspace{-1.2cm}
\caption{Process Deviations, Product Deviations, and Fixing Steps}
\label{fig:deviation}
\end{center}
\vspace{-0.6cm}
\end{figure}

\textit{Step 1: Determine Solution Model, Distance, and Optimal Problem Solving
Paths.} First, we create a model representing the correct solution (i.e.,
$S_{S}$) for the modeling task. Subjects had to work on a
re-modeling task as described in Sect. \ref{exploratoryStudy}, i.e., in addition to an informal textual description they
obtained the solution to the
modeling task in the form of a graphical model. Thus, the goal state of the modeling
task was clearly defined and unique, i.e., subjects should create a graphical
representation of the process that exactly looks like the solution model. To be
able to assess not only how closely subjects reached the goal state
(i.e., how similar their resulting model is to the solution model), but also how
efficiently their problem solving process was, we determine the
\textit{distance} for transforming an empty model $S_0$ to $S_{S}$,
i.e., the minimum number of change patterns required for the respective model transformation. Generally, there are
several options to create the solution model $S_{S}$ by starting from $S_0$ and
applying a sequence of model transformations. From a cognitive perspective, each
sequence of change patterns that leads to $S_{S}$ without detours constitutes an
\textit{optimal problem solving path}. Starting from $S_0$ the process fragment
depicted in \figurename~\ref{fig:deviation} can be created with  6 change
patterns; e.g., $S_{S}$ can be created by first inserting \textit{A} and then
\textit{B}, next embedding \textit{B} in a conditional branch, then updating the
two transition conditions, and finally inserting \textit{C} ($P_0$ in
Fig.~\ref{fig:deviation}).

\textit{Step 2: Determine Deviations from Solution Model and Optimal Problem Solving Path.} 

To quantify the efficiency of the problem
solving strategy used by the subjects to accomplish the re-modeling task, their problem solving path is analyzed.
To be more specific, using the replay feature of CEP we
compare the subject's problem solving path $P$ with the optimal one and capture deviations from it. For this, every superfluous change pattern application a subject performs is counted as a \emph{process deviation}.
To quantify how close subjects reached the goal state (i.e., how
 similar their resulting model is to the solution model $S_S$) we consider \emph{product deviations} that measure the
number of incorrect change pattern applications leading to deviations between the final models created by
the subjects and the solution model $S_S$.

Fig.~\ref{fig:deviation} shows the problem solving path $P_0$ of one modeler who managed to model the depicted fragment correctly (i.e., 0 process deviations and 0 product
deviations). Problem solving path $P_2$, in turn, leads to a correct goal state (i.e., 0 product
deviations). However, the modeler made a detour of 2 change patterns before reaching the
solution (i.e., solution path $P_2$ comprises 2 superfluous change patterns
summing up to 2 process deviations). Now assume that the modeler, who
took a detour when creating the process model, did not correct the introduced
error ending up with an incorrect process model (cf. path $P1$ in
Fig.~\ref{fig:deviation}). The application of the \textsf{Embed in Loop} pattern
(instead of \textsf{Embed in Conditional Branch}) constitutes 1 product
deviation (i.e., the modeler applied one incorrect change pattern that led to an incorrect goal state).

\noindent  Since not every subject reached the goal state (i.e.,
their models contain product deviations), the direct comparison of process
deviations might favor modelers that left out parts that were difficult to model
and where other subjects produced a high number of process deviations. To
decrease this bias we consider a second measure for operationalizing problem
solving efficiency. In addition to the process deviations described above this
measure considers the effort needed to  correct an incorrect process model
(denoted as \emph{fixing steps}), i.e., the steps needed to transform the
created model into $S_S$. For example, to correct the model that resulted from $P1$
in Fig.~\ref{fig:deviation}, 5 fixing steps are needed, irrespective of whether or not the core or the extended change pattern set is used. First the
fragment embedded in the loop has to be deleted. Next, \textit{B} has to be
re-inserted and embedded in a conditional branch, and then the two transition
conditions must be updated.
Fixing steps and process deviations are then combined in a single measure called
\emph{total process deviations}. This measure does not only consider detours
(i.e., process deviations), but also model transformations that would be needed
to correct the process model (i.e., resolving product deviations).

\textit{Step 3: Detection of Outliers.}
In order to limit the influence of modelers who are experiencing severe
difficulties during the creation of the process model, we test for outliers
regarding the number of process deviations. For this purpose, we utilize the
Median Absolute Deviation (MAD) to detect outliers. More
specifically, we apply a rather conservative criterion for removing outliers by
removing values that differ at least 3 times the MAD from the
median~\cite{Ley+13}. As a result, one PPM instance was removed from Group B
regarding further analysis.

\vspace{-0.2cm}
\subsection{Results}
\label{taskA}
\paragraph{Descriptive Statistics.}
To create a correct solution model 28 operations are needed. Overall, 123
process deviations (i.e., detours in the modeling process) and 44 product
deviations (i.e., deviations of the final models from the solution model) were
identified (cf. Table~\ref{tab:OverviewOfDeviations}). From the 123 process
deviations 60 can be attributed to Group A (3.53 deviations per subject), while
63 were found for Group B (3.94 deviations per subject). In terms of product
deviations they were equally distributed among the two groups, i.e., 22 product
deviations per group (1.29 deviations per subject in Group A and 1.38 deviations
per subject in Group B). In order to resolve the product deviations, 45 fixing
steps are required for the models of Group A and 29 fixing steps for Group B resulting in
105 and 92 total process deviations respectively.

\begin{table}
\centering
\begin{tabular}{lrmmm}
Scale & Group A & Group B \\
\hline
Process deviations & 60 & 63 \\
Average Process deviations per modeler & 3.53 & 3.94 \\
Product deviations & 22 & 22 \\
Average Product deviations per modeler & 1.29 & 1.38 \\
Fixing steps & 45 & 29 \\
Average fixing steps per modeler & 2.65 & 1.81 \\
Total process deviations & 105 & 92 \\
Average total process deviations per modeler & 6.18 & 5.75 \\
\hline
\end{tabular}
\caption{Overview of Deviations}
\label{tab:OverviewOfDeviations}
\vspace{-0.5cm}
\end{table}

\paragraph{Hypothesis Testing.}
We test for significant differences between the two groups regarding process
deviations and total process deviations. We expect that the modelers using the
extended pattern set have significantly fewer process deviations, because the
extended pattern set allows them to resolve detours with fewer steps.
Moreover, we expect an impact on the total process deviations, since the
extended pattern set allows transforming the model created by the modelers
with fewer steps into the solution model.

\textbf{Hypothesis H4} \textit{The usage of an extended change pattern set significantly decreases the number of process deviations.}

\textbf{Hypothesis H5} \textit{The usage of an extended change pattern set significantly decreases the number of total process deviations.}

To test for differences in terms of process and total process deviations,
we apply the t-test since the data was normally distributed. No statistical
difference could be observed for process deviations ($t(31)=-0.24, p=0.82$) or
total process deviations ($t(31)=0.25, p=0.81$). 

\paragraph{Discussion.}
Our results did not yield statistically significant differences between the two
groups. This indicates that the usage of an extended change pattern set might
not have an impact on both process deviations and total process
deviations. An alternative explanation could be that process modelers did not
use the provided patterns frequently enough to obtain statistically significant
differences (i.e., pattern adoption was low). Another explanation could be
that subjects did not use the patterns effectively, canceling out a potential
positive impact. To investigate these alternative explanations in more detail
Sect.~\ref{advantagesAndDisadvantages} analyzes the actual use of the move
change patterns.

\section{Actual and Potential Use of an Extended Pattern Set}
\label{advantagesAndDisadvantages}
This section addresses research question RQ3 which deals with the actual use of
the additional change patterns compared to the potential usage of those
patterns. The analysis of invocations of the move change patterns revealed that the serial
move pattern was only applied 3 times (by 3 different participants), whereas the
parallel move pattern was used 18 times (by 7 different participants). This
indicates that the subjects adopted the move patterns only to a limited extend.
Out of the 21 move pattern applications, 11 were correct; i.e.,
they led to correct intermediate models, either directly through the
application of the pattern or the application of the pattern in combination with
additional patterns. In turn, 10 applications of the parallel move
pattern were incorrect and either led to an undesired model or did not
yield any changes of the model. This indicates that subjects had difficulties
when applying the move change patterns.

Though the actual use of the move patterns was limited, we investigate their
theoretical potential. For this, we analyze the number of fixing steps
required to correct a model with product deviations (i.e., to transform it into
$S_S$). We further analyze how the availability of an extended
pattern set impacts this measure. In a second step, we analyze the potential
of an extended pattern set for reducing process deviations, i.e., by enabling a faster
resolution of detours.

\begin{table}
\centering
\begin{tabular}{lmm}
Scale & Group A & Group B \\
\hline
Fixing steps with move & 45 & 64 \\
Fixing steps without move &  25 & 29 \\ 
Saved operations & 20 & 35 \\
\hline
Process Deviations & 60 & 63 \\
Unnecessary Operations & - & 15 \\
Saved operations & 9 & 0 \\
Potential process deviations &  51 & 48 \\
\hline
\end{tabular}
\caption{Potential Use of the Move Change Pattern}
\label{tab:PotentialUse}
\end{table}

To show the potential of an extended pattern set for resolving product
deviations, Table~\ref{tab:PotentialUse} depicts the number of fixing steps,
when using the core pattern set and for the extended pattern set. For Group A,
45 fixing steps are required to correct all product deviations that occurred. By
making the extended pattern set available to Group A, this number could be
reduced to 25 (i.e., 20 fixing steps could be saved).
In turn, for Group B the number of observed fixing steps is 29.
Without the extended pattern set, however, 64 fixing steps would be needed. This
indicates the theoretical potential of the extended pattern set for reducing the
number of fixing steps and, thus, the number of total process deviations.

To investigate the potential for reducing process deviations for Group A, we
analyze whether process deviations could have been reduced when using move
change patterns. In turn, for Group B we focused on the number of operations
that would have been saved if move patterns were always applied
correctly. As illustrated in Table~\ref{tab:PotentialUse}, 9 operations could be
saved if the move pattern had been available for Group A resulting in 51
potential process deviations. Regarding Group B, 15 operations could have been
saved through correct pattern application resulting in 48 potential process
deviations.

% \begin{table}
% \centering
% \begin{tabular}{lrrrrr}
% Group & Process Deviations & Unnecessary Operations & Saved operations & Potential process deviations \\
% \hline
% A & 60 & - & 9 \\
% B & 63 & 15 & 0 \\
% 
% \hline
% \end{tabular}
% \caption{Potential for Reducing Process Deviations}
% \label{tab:PotentialProcess}
% \end{table}

\paragraph{Discussion.}
These results suggest that a theoretical potential for using move change
patterns exists. However, the subjects used the move change patterns only to a
limited extent and had troubles with their correct application.
As a consequence the potential of the additional patterns could not be fully
exploited. Since mental effort and perceived ease of use is lower with the core
pattern set it might be more favorable to use the core pattern set for
process modelers that are only moderately familiar with process modeling and are
no experts in the usage of change patterns. We might speculate that the extended
pattern set could be promising for more experienced users (who are literate in
pattern usage). 
\section{Limitations}
\label{discussion} 

As with every other research, this work is subject to several
limitations. Certainly, the relatively small sample size constitutes a threat
regarding the generalization of our results. Using students instead of
professionals poses another threat regarding external validity. In previous
research with software engineering students it has been shown that students may
provide an adequate model for the professional
population~\cite{DBLP:journals/ese/HostRW00,DBLP:journals/ese/PorterV98,Run03}.
Still, generalizations should be made with care.
Moreover, since we used subjects who were moderately familiar with process
modeling and change patterns results cannot be generalized to expert modelers.
It can be assumed that process modelers experienced with the usage of change
patterns will presumably face less problems during model creation and will be
able to apply patterns more effectively.
Another limitation relates to the fact that we used only one modeling task in
our study. The potential benefit of move patterns, however, depends on the
structure of the process model to be created. For more complex process models
with higher nesting depth the potential usefulness might be higher. Thus, it is
questionable in how far results may be generalized to models with different
characteristics. As a consequence, we plan further experiments testing the
impact of model structure on challenges regarding change pattern usage.
Moreover, this work compares two particular change pattern sets. Using an
extended change pattern set with different patterns (e.g., a pattern to change a
conditional fragment into a parallel fragment or to change a conditional
fragment to a loop) might lead to different results. Another limitation
regarding the external validity relates to the process modeling notation (i.e.,
BPMN) and the modeling tool used (i.e., CEP). Results might be different when
using other modeling languages or different modeling tools.

\section{Related Work}
\label{relatedWork}

% \subsection{Quality Frameworks and Process Model Quality}
% 
% Different frameworks and guidelines have been developed that define quality
% aspects in the context of process models. The SEQUAL framework uses semiotic
% theory for identifying various dimensions of process model quality~\cite{KrSJ06}.
% The Guidelines of Modeling (GoM) describes various quality considerations for
% process models~\cite{Bec+00} and prescribes principles such as correctness and
% clarity that should be taken into account. The so-called `Seven Process Modeling
% Guidelines' (7PMG) accumulate the insights from various empirical studies
% (e.g.,~\cite{MVDA+08,Mood05}), to develop set of actions a process modeler may
% want to undertake to avoid issues with respect to the understandability of a
% process model~\cite{MeRA10}. Similar metrics have been proposed to assess the
% quality of the model artifact itself (e.g., \cite{AgGP06,Card08,GrLa06,RCRP09}).
% Pragmatic quality (i.e., the process models understandability), has been
% investigated considering insights from cognitive psychology
% in~\cite{ZPM+11,ZuPW11,ZSPW12}. 

The presented work relates to research developed in the
 context of the creation of process models and process model creation patterns.
 
Research on the creation of process models builds on
observations of modeling practice and distills normative procedures for steering
the process of modeling toward successful completion. To do so,
~\cite{FrWe06,HoPW05} deal with  structured discussions among different parties
(system analysts, domain experts). In this line of research, \cite{Ritt07}
analyzes the procedure of developing process models in a team, while
\cite{StPS07} discusses participative modeling. Complementary to these works,
whose focus is on the effective interaction between the involved
stakeholders, our work focuses is on the \emph{formalization} of the
process model.

Researchers have also focused on the interactions with the modeling
environment, i.e., the PPM. \cite{PSZ+13} identified
three distinct modeling styles, whereas \cite{PZW+12,CVP+13} suggest different
visualization techniques for obtaining an overview of the PPM; \cite{CVR+12} demonstrates
that a structured modeling style leads to models of better quality.
\cite{PFM+13} investigates the PPM using eye movement analysis. While these
works focus on interactions with the modeling environment based on change
primitives, this paper investigates the use of change patterns.

Change patterns for process model creation have been investigated as well; e.g.,
AristaFlow allows modeling a sound process schema based on an extensible set of
change patterns~\cite{DaRe09}. \cite{Gschwind08} describes a set of pattern
compounds, comparable to change patterns, allowing for the context-sensitive
selection and composition of workflow patterns. Complementary to these works,
which have a strong design focus, this paper provides empirical insights into
the usage of change patterns. More precisely, it builds upon the results
obtained in \cite{WPVR13}, which describes recurring challenges modelers face
during the PPM using change patterns.

\section{Summary}
\label{summary} 
While recent research has contributed to a better understanding
 regarding the PPM, little is known about this process when utilizing change
 patterns. In
this experiment we investigate the impact of the available patterns on the PPM
and the modeler's perception.
The results indicate that an extended change pattern set puts an additional
burden on modelers who perceive them as more difficult to use. In addition, when
using these patterns, subjects faced considerable difficulties.
Therefore, (against our expectations) our data does not indicate an increased
problem solving efficiency, i.e., the expected benefits of using the extended
change pattern set did not materialize. This indicates that the change pattern
set should be selected with care, especially for modelers with limited
experience. Future research should include investigations on new change pattern
sets having a (theoretical) potential for reducing process deviations, e.g., a
pattern to change a conditional fragment into a parallel or a loop fragment.

\bibliographystyle{splncs}
\bibliography{literature}

\end{document}